\documentclass[useAMS,usenatbib]{mn2e}
\usepackage{amssymb,amsmath,epsfig,times,natbib,longtable,color,graphicx}

\long\def\symbolfootnote[#1]#2{\begingroup%
\def\thefootnote{\fnsymbol{footnote}}\footnote[#1]{#2}\endgroup} 

\newcommand{\xmm}{{\it XMM-Newton}}

\makeatletter
 \def\hlinewd#1{%
   \noalign{\ifnum0=`}\fi\hrule \@height #1 \futurelet
    \reserved@a\@xhline}
\makeatother

\title[UFO variability]{Revealing the ultra-fast outflow in IRAS~13224-3809 through spectral variability}

\author[M. L. Parker et al.]{M. L. Parker$^{1}$,\thanks{Email: mlparker@ast.cam.ac.uk}
W. N. Alston$^{1}$
D. J. K. Buisson$^{1}$,
A. C. Fabian$^{1}$,
J. Jiang$^{1}$,
E. Kara$^2$,\newauthor
A. Lohfink$^{1}$,
C. Pinto$^{1}$
and
C. S. Reynolds$^{2}$\\
  $^1$Institute of Astronomy, Madingley Road, Cambridge, CB3 0HA \\
  $^2$Department of Astronomy, University of Maryland, College Park, Maryland 20742-2421, USA
}
\date{}

\begin{document}

\maketitle

\begin{abstract}
We present an analysis of the long-term X-ray variability of the extreme narrow-line Seyfert 1 (NLS1) galaxy IRAS~13224-3809 using principal component analysis (PCA) and fractional excess variability ($F_\mathrm{var}$) spectra to identify model-independent spectral components. We identify a series of variability peaks in both the first PCA component and $F_\mathrm{var}$ spectrum which correspond to the strongest predicted absorption lines from the ultra-fast outflow (UFO) discovered by \citet{Parker17_iras}. We also find higher order PCA components, which correspond to variability of the soft excess and reflection features. The subtle differences between RMS and PCA results argue that the observed flux-dependence of the absorption is due to increased ionization of the gas, rather than changes in column density or covering fraction. This result demonstrates that we can detect outflows from variability alone, and that variability studies of UFOs are an extremely promising avenue for future research.
\end{abstract}


\section{Introduction}

Ultra-fast outflows (UFOs) are defined as the subset of X-ray detected outflows with an observed velocity of $>10000$~km~s$^{-1}$, with velocities extending up to $\gtrsim0.3c$ \citep[e.g.][]{Tombesi10}. These outflows are an exciting area of research \citep{Tombesi15, Nardini15}, as they may be partly responsible for driving AGN feedback \citep[see review by][]{Fabian12_feedback}.
The most common technique for detecting such outflows is searching for blueshifted Fe\textsc{xxv} or Fe\textsc{xxvi} features in the hard X-ray band (7--10~keV), frequently using blind line scans \citep{Cappi09,Tombesi10, Gofford13}. While there is less signal in this band, these features are the strongest at high ionizations, and this band is not confused by the presence of lower ionization features from warm absorbers or Galactic absorption. 

A major advantage of studying compact objects is the rapid variability they display, which gives us another window to understanding their complex behaviour. X-ray reverberation mapping, for example, has the potential to reveal the geometry of the inner accretion disk around the black hole \citep{Uttley14}. While there are a large number of powerful techniques for analysing the variability properties of AGN they have rarely been applied to UFOs. If these techniques can be applied to such outflows, we can hope to understand their behaviour in far greater detail.

In recent work, we have greatly expanded the use of principal component analysis (PCA) as a tool for understanding the variability of AGN \citep{Parker14_mcg6, Parker14_ngc1365, Parker15_pcasample}. However, a persistent weakness of this technique is its insensitivity to narrow features, due to the low fraction of the total (i.e. broad-band) spectral variability they contribute, relative to the continuum. Because the variability fraction is used to determine which components are produced by a real signal and which by noise, the low variability of narrow spectral features makes them indistinguishable from noise in many cases.
A key strength of PCA as a tool for studying AGN variability is that it allows us to quickly and easily make qualitative comparisons of the variability in different sources, and scales trivially to arbitrarily large datasets. 
Relative to the fractional root-mean-square (RMS) variability amplitude \citep[$F_\mathrm{var}$,][]{Vaughan03_variability}, the main advantage of PCA is that it returns additional, higher order variability terms, corresponding to different spectral components, rather than just the sum of all variability.
If the difficulties in studying narrow features can be overcome, PCA may have several useful applications in studying UFOs, particularly with large datasets and slow variability.

\begin{figure*}
\centering
\includegraphics[height=8cm]{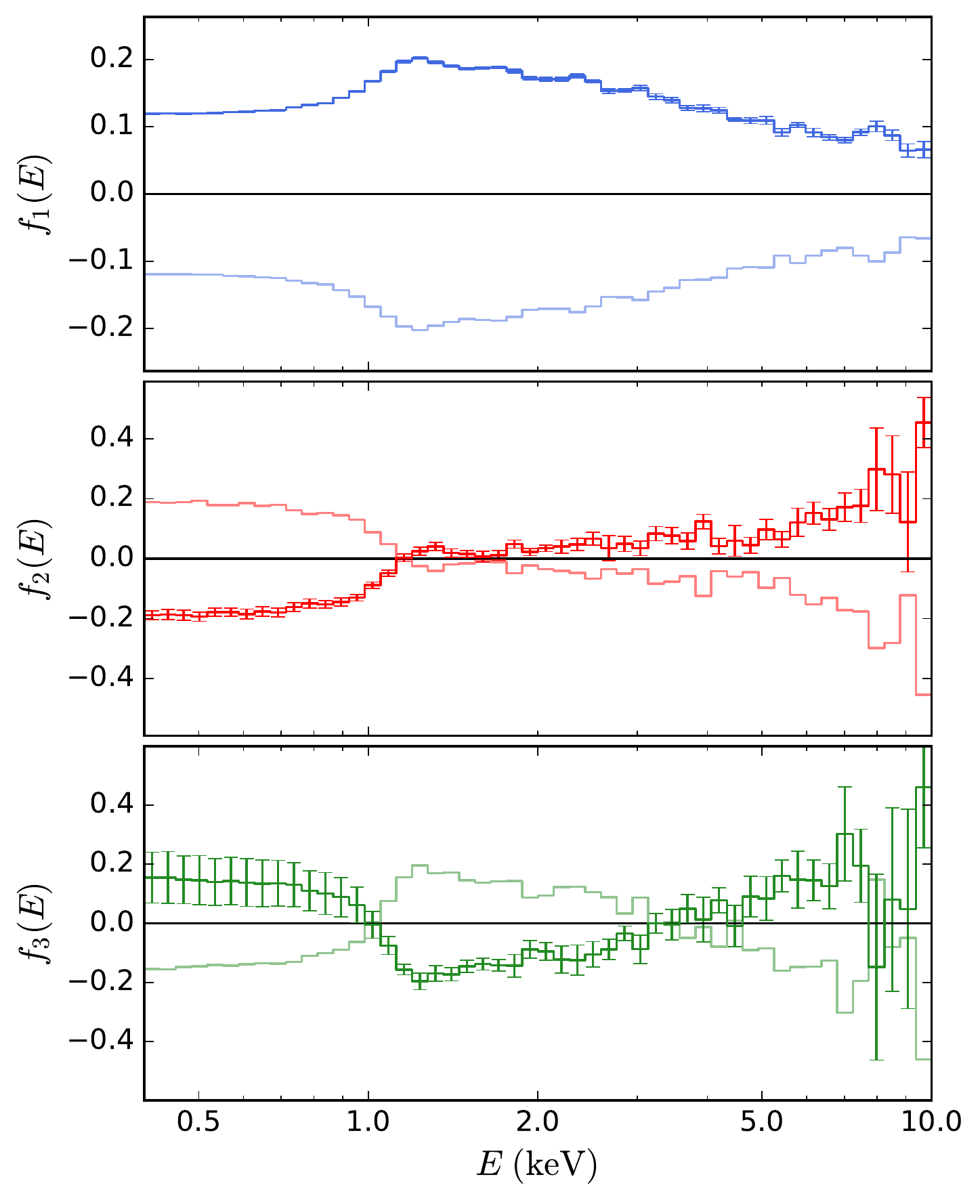}
\includegraphics[height=8cm]{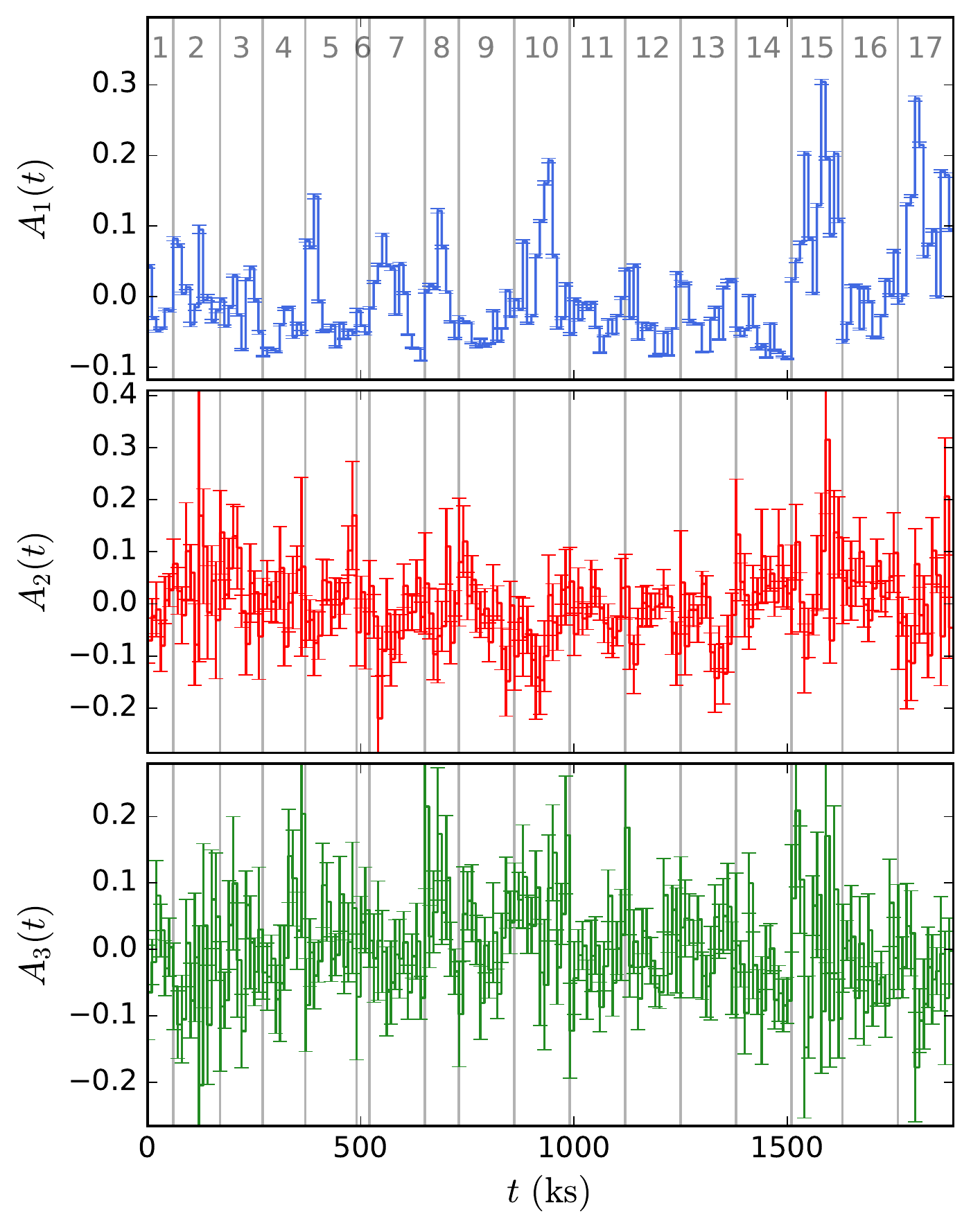}
\caption{Left: Component spectra for the first three PCs from the analysis of all \xmm\ EPIC-pn data of IRAS~13224-3809. We show mirror images of the spectra, as the sign of the y-axis is essentially arbitrary. Energies are in the observer's frame. Right: Corresponding lightcurves, showing the contribution of each component to each 10~ks spectrum}. Vertical lines and numbers at the top correspond to the different observations.
\label{fig_epicpn_pcs}
\end{figure*}

In this paper, we examine the recently discovered UFO in IRAS~13224-3809 \citep[hereafter P17]{Parker17_iras}, the most x-ray variable AGN, using PCA. IRAS~13224 has very strong reflection signatures \citep{Fabian13_iras, Chiang15}, produced by relativistic blurring of fluorescent emission, scattered off the inner accretion disk. \citet{Kara13_iras} demonstrated that this reflection produces a lag between the direct continuum and the reflected emission, corresponding to the difference in light travel-times. This lag appears to be flux dependent, possibly due to changes in the geometry of the X-ray corona. IRAS~13224-3809 was the subject of an extremely large (1.5Ms) observing campaign with \xmm\ in 2016, aiming to explore these lags in great detail and use them to map out the inner accretion disk. Unexpectedly, in P17, we detected multiple absorption features with a combined significance of $>7\sigma$ from a UFO at 0.236$c$. We demonstrated that this UFO is extremely variable on timescales of hours or less, and that it responds rapidly to changes in the continuum flux. This rapid and extreme variability makes this UFO the best candidate for analysis with timing techniques, which have the potential to greatly expand our knowledge of the physics of AGN outflows.

The paper is organized as follows:
\begin{itemize}
\item In \S~2.1 and 2.2, we describe the data and the PCA process.
\item In \S~3, we present the results of our analysis, with simulation results in \S~3.1
\item In \S~4, we discuss the implications of this result, and present our conclusions in \S~5.
\end{itemize}

\section{Observations and Data Reduction}
\label{section_datareduction}

\subsection{X-ray observations}
\label{section_xray_observations}

The \xmm\ data are reduced as in P17. We use only the EPIC-pn data, which are extracted using the \textsc{epproc} ftool. We filter the data for background flares, and extract source and background spectra from 40$^{\prime\prime}$ circular regions, avoiding the area of the pn detector with a high copper background.

\subsection{PCA}
We follow the PCA method described in \citet{Parker14_mcg6}, using singular value decomposition (SVD) to decompose a matrix of spectra into a set of principal components (PCs) which account for the majority of the variability of the source.

We slice the data into 10~ks intervals from the start of each observation, extracting spectra for each interval. We then construct a grid of spectra $F(E,t)$ (or lightcurves, if transposed) with columns corresponding to energy, and rows corresponding to time. This is then transformed to normalized residual spectra:
\begin{equation}
f(E,t)=\frac{F(E,t)-\bar{F}(E)}{\bar{F}(E)}
\end{equation}
where $\bar{F}(E)$ is the time-averaged spectrum. PCA then decomposes this grid into a set of spectral components $f_{i}(E)$ and normalizations $A_{i}(t)$ such that
\begin{equation}
f(E,t)=\sum\limits_{i=1}^{n} f_i(E)\times A_i(t)
\end{equation}
The amplitude of the components $f_i$ can be understood as the variance spectra $\sigma^2(E)$ of each spectral component, and the sign on each bin indicates whether it is correlated or anti-correlated with the other bins. SVD returns three outputs: a matrix of component spectra, a set of eigenvalues corresponding to the total variance of each component, and a matrix of lightcurves for each component\footnote{In \citet{Parker14_mcg6} the lead author forgot the existence of these lightcurves, and instead fit the output components back to the data to achieve the same effect. We do not recommend this approach.}.
We calculate the errors following the Monte-Carlo approach of \citet{Miller08} by perturbing the input spectra with Poisson noise and repeating the analysis, then measuring the standard deviation in the results.

\section{Results}

We find three strongly significant principal components using the log-eigenvalue (LEV) plot \citep[see e.g.][]{Parker14_mcg6}. This is a simple diagnostic tool for identifying which components are likely to be significant - components due to noise should follow a geometric decay, making a straight line on the log plot. We find the first three components to be strongly significant. PCs 4 and 5 may be marginally significant, but visual inspection of their spectra shows them to correspond to single bins at high energies, most likely due to (or indistinguishable from) noise. We therefore restrict the analysis to the first three components.

\begin{figure*}
\includegraphics[width=0.7\linewidth]{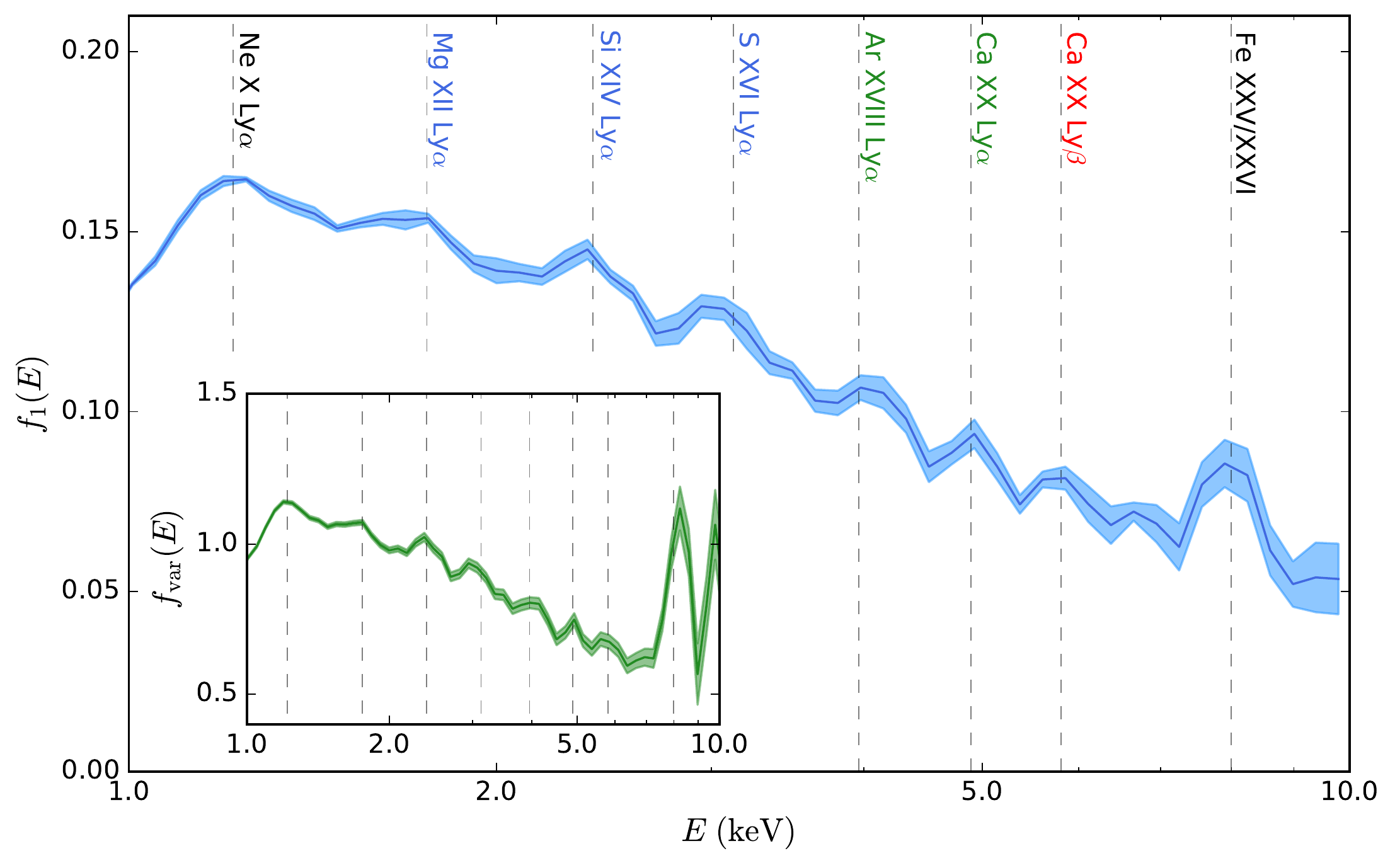}
\caption{Zoom of PC1 from our analysis, from 1--10~keV, with the energies of the strongest UFO lines predicted by the model from P17 marked by dashed lines. Previously detected lines from P17 are in black, new features that we can spectroscopically confirm are blue, new features predicted by simulations are green, and the rather implausible Ca\textsc{XX} Ly$\beta$ (which is not predicted or found in the spectra) is shown in red. Inset shows the RMS spectrum calculated from the same data, with the same lines marked. Energies are in the observer's frame. }
\label{fig_PC1_big}
\end{figure*}

Fig.~\ref{fig_epicpn_pcs} shows the spectra of the first three PCs and their respective lightcurves. A few features are immediately obvious:
\begin{itemize}
\item PC1 is fairly flat, and all positive, with either a dip at 7~keV or a peak at 8~keV, and suppressed below $\sim1$~keV, where the soft excess lies. This general shape is not uncommon in the sources we have previously examined, and corresponds well to simulations of a variable power-law continuum \citep{Parker15_pcasample}.
\item PC2 shows pivoting behaviour, with an anti-correlation between low and high energies. It also shows a soft excess. Pivoting behaviour is common in other AGN, with much of the sample in \citet{Parker15_pcasample} showing similar pivoting.
\item PC3 shows an anti-correlation between intermediate energies ($\sim1$--3~keV) and low and high energies, with an peak at $\sim7.5$~keV (the peak of the broad iron line) and a broad soft excess. This is frequently seen in sources with a strong reflection component, where the broad line and soft excess are correlated. 
\end{itemize}

\begin{figure*}
\centering
\includegraphics[width=0.7\linewidth]{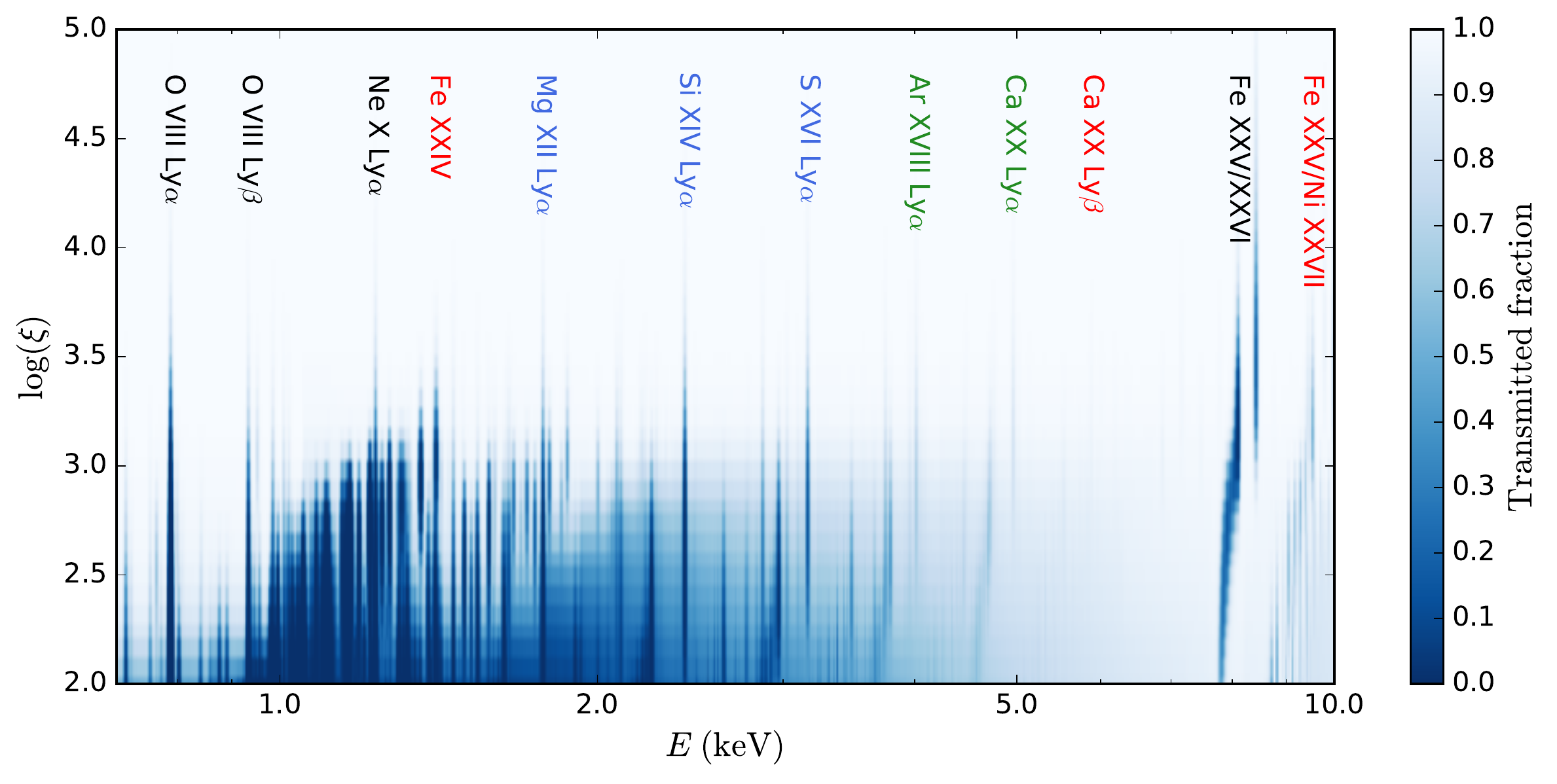}
\caption{Map of the strongest absorption features as a function of energy and ionization, calculated using \textsc{warmabs} and blueshifted to match the observer's frame for IRAS~13224-3809. The strongest lines are marked, with previously detected lines in black, new spectrally confirmed features in blue, new features predicted by the model in green, and features predicted and not found or found and not predicted in red. Energies are in the observer's frame.}
\label{fig_linemap}
\end{figure*}

Closer examination of PC1 reveals additional structure. We show an expanded version of this component above 1~keV in Fig.~\ref{fig_PC1_big} (below 1~keV the variability of this component is suppressed and the resolution of the EPIC-pn is poor, so no features are detected). Strikingly, there are a series of peaks in the variability, corresponding to the predicted energies of the strongest absorption features in the physical model of the UFO from \citet{Parker17_iras}. We also calculate the $F_\mathrm{var}$/RMS spectrum, plotted on the inset of this figure. The RMS spectrum shows the same series of peaks, with a significantly larger feature at the Fe\textsc{xxv} line energy and a variability spike at the edge of the bandpass.
For reference, we show a map of the strongest lines predicted by \textsc{warmabs} in Fig.~\ref{fig_linemap} (note that some of these features are a blend of multiple lines, in which case we name the line with the highest equivalent width over the ionization range of interest). Features corresponding to 8 of these lines appear in the variability spectra, and two more were detected in the RGS data presented in P17.

\begin{figure}
\includegraphics[width=\linewidth]{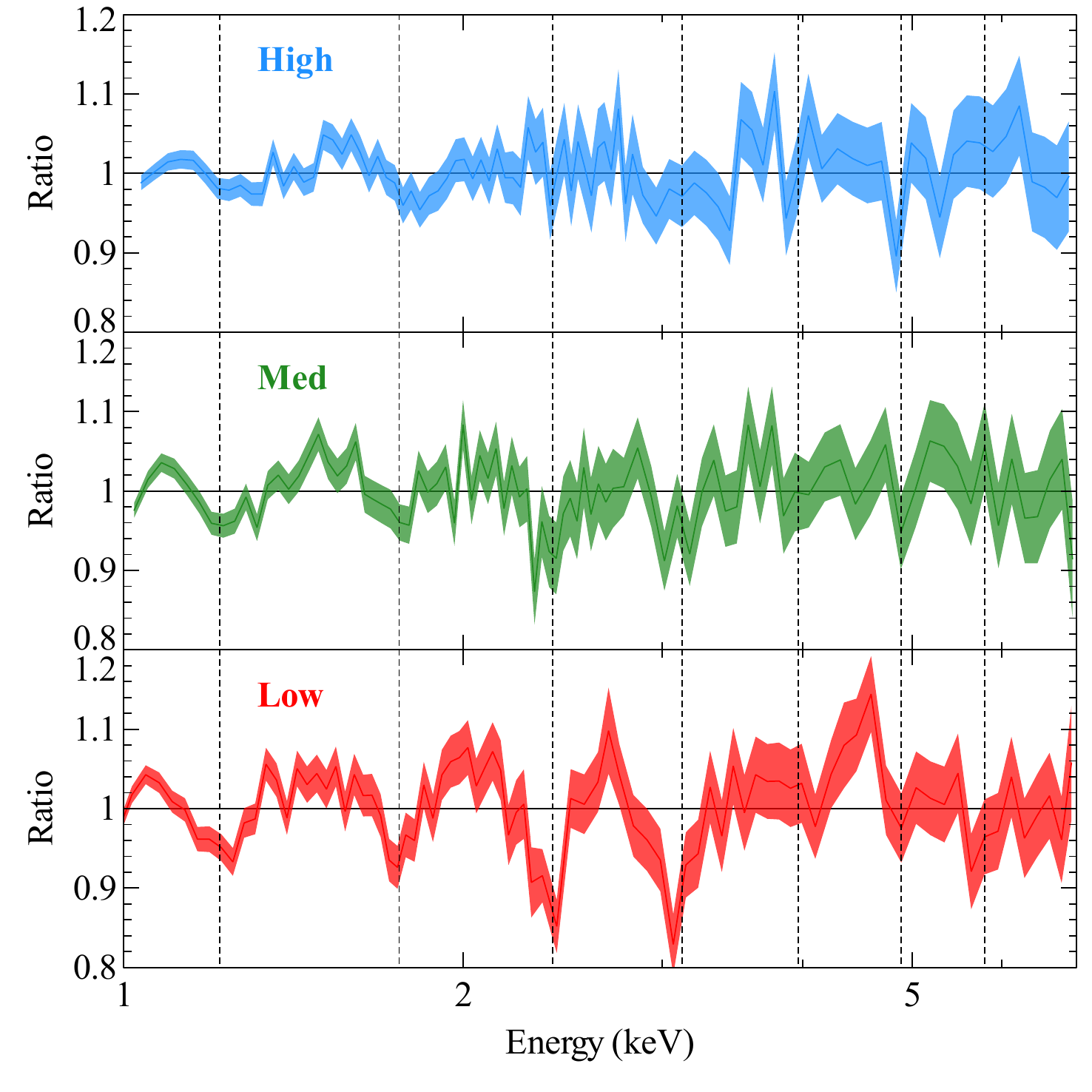}
\caption{Ratios of the three flux resolved spectra from P17 to a simple phenomenological model, showing strong, flux-dependent absorption features at the energies where we find excess variability and where predicted by the UFO model. Energies are in the observer's frame.}
\label{fig_lineratios}
\end{figure}

We perform a simple visual inspection to look for these additional lines in the time-averaged spectra. We fit the three flux states from P17 between 1 and 7 keV with a simple phenomenological model: two power-laws plus a black body. The purpose of this phenomenological model is to allow narrow residual features to be seen against the smooth continuum. The model does a reasonable job of describing the soft-excess plus power-law continuum, and the second power-law accounts for the additional curvature due to the presence of the broad iron line in the reflection spectrum (as the line peaks at 7keV). The results are shown in Fig.~\ref{fig_lineratios}. In the low flux spectrum absorption features due to the Ne/Fe blend, Mg XII, Si XIV and S XVI are all clearly visible, and disappear from the spectrum as the flux rises.

\subsection{Simulated Variability}

To further investigate the UFO variability seen with \xmm , we perform a simple simulation as in \citet{Parker15_pcasample}: by simulating a set of 50 fake spectra based on a simplified model and analysing them with PCA (and calculating the RMS spectrum), we can compare to the results with real data and identify the likely causes of the observed variability. To produce qualitatively similar variability, we use a simple power-law continuum with an ionized outflow modelled using \textsc{warmabs}, with the same net blueshift as our data. We vary the normalization of the power-law, $N_\mathrm{pl}$, randomly between 1 and 4, and set the ionization of the absorber to be
\begin{equation}
\log(\xi) = \log_{10}(N_\mathrm{pl})+3.5
\end{equation}
This gives a range of $\log(\xi)$ from 3.5--4.1. This is likely smaller than the range expected from the extreme variability of IRAS~13224-3809, where the flux changes by a factor of $\sim50$ or more, but it gives a conservative baseline model from which to extrapolate the variability behaviour without introducing potential additional complicating factors. This model is essentially a minimum working example for the UFO variability: the continuum is as simple as possible, to isolate the expected features from an outflow where the ionization responds to the continuum.

\begin{figure}
\centering
\includegraphics[width=\linewidth]{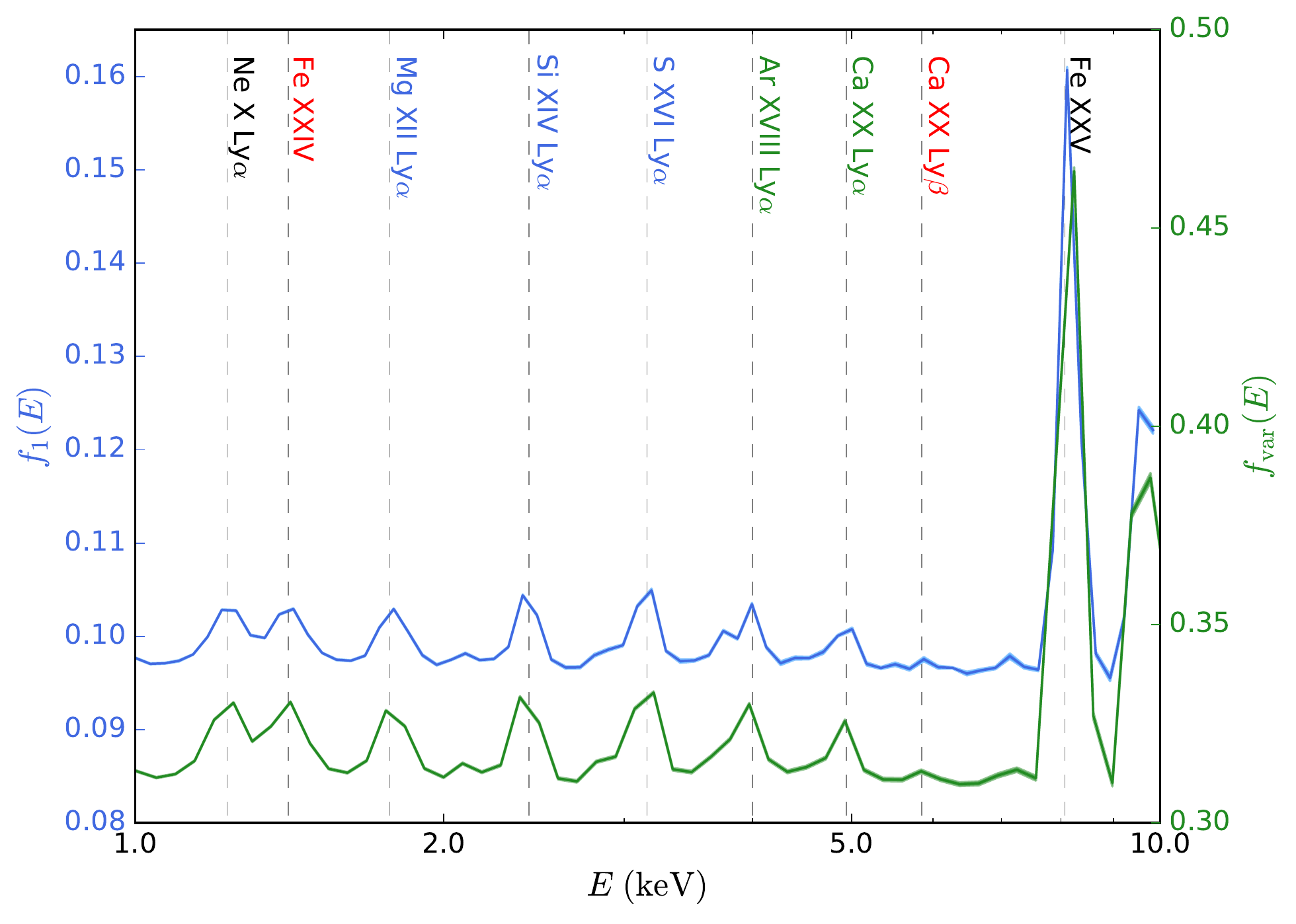}
\caption{Simulated PCA (blue, top) and RMS (green, bottom) spectra for a simple UFO$\times$power-law model. Over this range of ionizations, the two are equivalent. Energies are in the observer's frame. Labels are coloured as in Fig.~\ref{fig_linemap}.}
\label{fig_sims}
\end{figure}

The result of this simulation is shown in Fig.~\ref{fig_sims}. Both PCA and RMS spectra show the same variability pattern, with a series of peaks at the same energies as we observe. The continuum is obviously different, and is due to us neglecting to include a less-variable component in addition to the power-law \citep[see simulations in][]{Parker15_pcasample}. Excluding the continuum shape, the main differences to the observed data are the relatively weak Ca\textsc{xx} Ly$\beta$ line and the relatively strong Fe\textsc{xxiv} and \textsc{xxv} lines. These differences are discussed further in \S~\ref{section_discussion}.

\section{Discussion}
\label{section_discussion}

The most immediately obvious aspect of our results is the detection of the 8~keV Fe\textsc{XXV} absorption line in the principal component and RMS spectra. The flux at the energy of the UFO line is significantly more variable than the continuum ($F_\mathrm{var}=1.1$, compared with 0.6 immediately before the iron line), and also correlated with the continuum flux. This can be caused by the UFO line responding to the continuum, as demonstrated in P17. This is most likely caused by the increased X-ray emission increasing the ionization of the wind. Building on this result are a series of seven variability peaks at the energies of the strongest absorption features predicted by the absorption models from P17, again found in both PC and RMS spectra. The energies of the lines we detect with PCA are a perfect match to the energies of the predicted lines, demonstrating that we are detecting the same absorption with spectroscopy and variability.
Visual examination of the spectra fit with a simple phenomenological model demonstrates that at least four of these lines (including the Ne/Fe blend just above 1~keV found in the RGS data in P17) are clearly present and strongly variable\footnote{A full statistical analysis of these features will be presented in future work (Jiang et al., in prep), as the final significance of the lines depends on the broad-band model used.}.

The broad-band shape of PC1 is similar to the strong reflection sources in \citet{Parker15_pcasample}, with a clear break around 1~keV, suppression of the variability at the energy of the soft excess, and a decline in variability with energy at high energies. This is simply explained by variability of the powerlaw continuum, which is strongest in the bands where the power-law is dominant and reflection is weakest (i.e. from 1--5~keV).
The reduced variability of PC1 at the soft excess rules out the possibility of the soft excess being due to absorption \citep[e.g.][]{Middleton07, Parker14_mrk1048}. Constant absorption has no effect fractional deviations, and thus no effect on the shape of the principal components, meaning that the suppression of PC1 at low energies must be due to an additive component that is less variable than the primary continuum.

The spectral shape of PC3 is the same as found in several other sources with strong reflection components, such as MCG--6-30-15 \citep{Parker14_mcg6}, and 1H~0707-495 \citep{Parker15_pcasample}. This component shows a correlation between the soft excess and iron line emission, with a peak at 7~keV where the relativistic iron line peaks. This spectral shape is characteristic of changes in the reflection fraction \citep[see simulations in ][]{Parker14_ngc1365, Parker15_pcasample}. Because of the strong reflection in this source \citep{Fabian13_iras}, this is likely the correct interpretation of this component.

There is an anti-correlation between soft and hard energy bins in the spectrum of PC2. The spectrum is relatively featureless, but with a strong soft excess. This may be an indication that there is an additional component to the soft excess as well as ionized reflection. There are several sources with a 2nd order pivoting term in the sample in \citet{Parker15_pcasample}, which can generally be explained by changes in the index of the power-law continuum. In this case, the soft excess in this component is very clear. This suggests that this component is responsible for some variability of the soft excess which is independent of both the primary continuum (PC1) and relativistic reflection (PC3). It is hard to be sure of the exact nature of this component from variability alone. It could represent a second process contributing to the soft excess, such as Comptonization, or it could be due to a second, higher ionization, reflection component, as found by previous studies of this source \citep{Fabian13, Chiang15}. We will discuss this more in future work, where we will explore the nature of the soft excess with several different methods.

One aspect to consider in this and in future work on UFO variability is the potential effect of emission from the wind. Photoionized emission lines from the wind might reasonably be expected to be less variable than the continuum, due to the smearing out of the variability caused by the wide range of light travel times. Such features would then appear in variability spectra as a series of dips at the corresponding energies. This is unlikely to be the dominant factor here, as the absorption lines are clearly much stronger than any corresponding emission, but it may be a factor in other sources.

Interestingly, the Fe\textsc{XXV} absorption feature is weaker in the PCA spectrum than the RMS spectrum and that predicted by simulations (which predict the same strength feature with both methods). Additionally, our simple simulation predicts a spike in variability at the edge of the bandpass, due to the complex of Fe/Ni lines at these energies. A possible such peak is found in the RMS, but not the PCA.  The reason for these differences is not immediately obvious, as generally PC1 and RMS spectra are a very good match to each other. Differences between the two must be caused by the slight difference in what they show: the RMS spectrum shows the total amplitude of the variability, whereas the PCA spectrum shows the amplitude of the \emph{correlated} variability. This therefore implies that the correlation between the Fe\textsc{xxv} line and the continuum is weaker than that of the other lines. This can be explained if the range of ionizations covered is large enough that the equivalent width of the Fe\textsc{xxv} line increases with flux at the lowest flux values, whereas the other lines (which are strongest at lower ionizations) are more uniform in their behaviour with respect to the continuum.
We therefore suggest a comparison between RMS and PCA spectra may be a useful diagnostic for understanding the complex behaviour of UFOs, as the subtle differences between the two contain a significant amount of information. The difference observed here is a strong argument for the ionization interpretation of these changes, as a change in column density or covering fraction of the absorber would produce the same response from all the lines, and hence no difference between the RMS and PCA spectra.

Of the line features found in Fig.~\ref{fig_PC1_big}, the Ar~\textsc{xviii} and Ca\textsc{xx} lines have no obvious spectral counterpart in the data shown in Fig.~\ref{fig_lineratios}. However, both the Ar~\textsc{xviii} and Ca\textsc{xx} Ly$\alpha$ lines are predicted  by our simulation at similar strengths to the other, spectroscopically confirmed, features, so we can be reasonably confident that they represent genuine absorption lines. This leaves just the Ca\textsc{xx} Ly$\beta$ line, which is very weak (so weak in \textsc{warmabs} that it is not visible in Fig.~\ref{fig_linemap}), not found in the spectra, and not predicted by the simulation. The best hope for determining whether this feature is real (which we consider extremely unlikely) is a detailed examination of flux dependent spectra with a full physical broad-band model, which will be addressed in future work. Interestingly, our simulation also predicts an Fe\textsc{XXIV} feature at 1.4~keV, where no such feature is observed in either variability spectrum. This could be as a result of the smaller ionization range considered by the simulation relative to the data, which may place more emphasis on lines which are strongly variable in this small range. We also do not detect the O\textsc{viii} lines at low energies found using the reflection grating spectrometer (RGS) in P17. The most likely reason for this is that these lines are simply too hard to detect with the much lower resolution of the EPIC-pn, which declines strongly in energy resolution at low energies.

Of the sources with UFO detections to date, IRAS~13224-3809 is probably the most suitable for analysis with PCA/RMS spectra, as it is the most X-ray variable AGN and also one of the most extensively observed. However, the method should also be suitable for application to many other sources. Because we do not require continuous observations, it can be applied to archival observations with arbitrary separations. In higher mass, less rapidly variable AGN this is a great benefit, as there is very little variability within a single observation.
The most obvious features for detecting a UFO are, as in spectral fitting, the high energy Fe\textsc{xxv/xxvi} lines. However, if the UFO ionization is low enough, this method can also pick up lower energy features (most likely the Si\textsc{xiv} and S\textsc{xvi} lines), without needing detailed spectral fitting or any additional computation. This is particularly useful for determining the ionization as a single Fe line cannot usually be uniquely identified with a single transition, leaving the velocity of the outflow ambiguous.
We recommend that the RMS spectra be used for initial searches, as it can return stronger features than PCA when the correlation between the UFO line strength and the continuum flux is non-trivial. PCA can then be compared with the RMS results to investigate the relationship between the source flux and the absorber, and to remove variability from other, uncorrelated, spectral components.

We note that large, unevenly sampled, multi-epoch datasets like this one are inherently difficult to analyse using Fourier-based techniques. This may be the reason why such features have not been discovered before, given the (well-deserved) popularity of advanced Fourier methods for understanding AGN variability, particularly time lags. If frequency domain techniques can be applied to UFO variability there may be a great deal of additional information we can extract (and long, almost continuous observing campaigns like the 2016 IRAS~13224-3809 observations are ideal for this).
IRAS~13224-3809 has previously been analysed using Fourier-based techniques to examine the reverberation lags \citep{Kara13, Chainakun16}. However, these studies are strongly targeted, focussing on specific energy bands and frequencies, with a coarse energy binning suitable for studying relativistic reflection. As such, we would not expect the UFO features to be obvious in these studies, but it may be possible to target them more towards outflows.

\section{Conclusions}
\label{section_conclusions}
We use PCA and $F_\mathrm{var}$ spectra to analyse the spectral variability of the extreme AGN IRAS~13224-3809. The results of this study are encouraging:

\begin{itemize}
\item We find clear signatures of the ultra-fast outflow using variability spectra, confirming the anti-correlation between the equivalent width of the absorption lines and the continuum flux. 
\item We find evidence for 7 different UFO absorption lines, with the same velocity as found in \citet{Parker17_iras} and predicted by our physical modelling of the outflow. At least five of these lines (including the Fe\textsc{xxv} line) are also clearly visible and flux-dependent in the spectra. Including the two O\textsc{viii} lines from P17, this brings the total number of separate lines detected from this UFO to 9.
\item By examining the difference between the PCA and RMS results, we argue that the change in absorption strength in IRAS~13224-3809 must be due to changes in the ionization, as changes in column density or covering fraction could not produce the observed differences between the two methods.
\item We find a third-order principal component with correlated soft excess and iron-line emission that corresponds to the strong relativistic reflection present in the source spectrum. The second-order component is more ambiguous, but most likely corresponds to an additional process contributing to the variability of the soft excess.
\end{itemize}

The ease with which we found these UFO lines using their variability suggests that this may be a powerful method for detecting outflows in low-mass, rapidly variable AGN, or less variable sources with multiple widely spaced observations. Variability studies represent the next logical step for understanding the behaviour of UFOs, potentially greatly expanding our knowledge of their location, density, and power.

\section*{Acknowledgements}
We thank the anonymous referee for detailed and constructive comments. MLP, CP, ACF, and AL acknowledge support from the European Research Council through Advanced Grant on Feedback 340492. 
WNA acknowledges support from the European Union Seventh Framework Programme (FP7/2013-2017) under grant agreement n.312789, StrongGravity. 
DJKB acknowledges support from the Science and Technology Facilities Council (STFC). 
JJ is supported by the Chinese Scholar Council (CSC) and Cambridge Trust Joint Scholarship Program.
CSR thanks NASA for support under the Astrophysical Data Analysis Program (ADAP) grant NNX17AF29G.
This work is based on observations with XMM-Newton, an ESA science mission with instruments and contributions directly funded by ESA Member States and NASA.

\bibliographystyle{mn2e}
\bibliography{bibliography_pds456}

\begin{thebibliography}{21}
\expandafter\ifx\csname natexlab\endcsname\relax\def\natexlab#1{#1}\fi

\bibitem[{{Cappi} {et~al}\mbox{.}(2009){Cappi}, {Tombesi}, {Bianchi}, {Dadina},
  {Giustini}, {Malaguti}, {Maraschi}, {Palumbo}, {Petrucci}, {Ponti},
  {Vignali}, \& {Yaqoob}}]{Cappi09}
{Cappi} M. {et~al.}, 2009, \aap, 504, 401

\bibitem[{{Chainakun}, {Young} \& {Kara}(2016){Chainakun}, {Young}, \&
  {Kara}}]{Chainakun16}
{Chainakun} P., {Young} A.~J., {Kara} E., 2016, \mnras, 460, 3076

\bibitem[{{Chiang} {et~al}\mbox{.}(2015){Chiang}, {Walton}, {Fabian},
  {Wilkins}, \& {Gallo}}]{Chiang15}
{Chiang} C.-Y., {Walton} D.~J., {Fabian} A.~C., {Wilkins} D.~R., {Gallo} L.~C.,
  2015, \mnras, 446, 759

\bibitem[{{Fabian}(2012)}]{Fabian12_feedback}
{Fabian} A.~C., 2012, \araa, 50, 455

\bibitem[{{Fabian}(2013)}]{Fabian13}
{Fabian} A.~C., 2013, in IAU Symposium, Vol. 290, IAU Symposium, {Zhang} C.~M.,
  {Belloni} T., {M{\'e}ndez} M., {Zhang} S.~N., eds., pp. 3--12

\bibitem[{{Fabian} {et~al}\mbox{.}(2013){Fabian}, {Kara}, {Walton}, {Wilkins},
  {Ross}, {Lozanov}, {Uttley}, {Gallo}, {Zoghbi}, {Miniutti}, {Boller},
  {Brandt}, {Cackett}, {Chiang}, {Dwelly}, {Malzac}, {Miller}, {Nardini},
  {Ponti}, {Reis}, {Reynolds}, {Steiner}, {Tanaka}, \& {Young}}]{Fabian13_iras}
{Fabian} A.~C. {et~al.}, 2013, \mnras, 429, 2917

\bibitem[{{Gofford} {et~al}\mbox{.}(2013){Gofford}, {Reeves}, {Tombesi},
  {Braito}, {Turner}, {Miller}, \& {Cappi}}]{Gofford13}
{Gofford} J., {Reeves} J.~N., {Tombesi} F., {Braito} V., {Turner} T.~J.,
  {Miller} L., {Cappi} M., 2013, \mnras, 430, 60

\bibitem[{{Kara} {et~al}\mbox{.}(2013{\natexlab{a}}){Kara}, {Fabian},
  {Cackett}, {Miniutti}, \& {Uttley}}]{Kara13_iras}
{Kara} E., {Fabian} A.~C., {Cackett} E.~M., {Miniutti} G., {Uttley} P.,
  2013{\natexlab{a}}, \mnras, 430, 1408

\bibitem[{{Kara} {et~al}\mbox{.}(2013{\natexlab{b}}){Kara}, {Fabian},
  {Cackett}, {Uttley}, {Wilkins}, \& {Zoghbi}}]{Kara13}
{Kara} E., {Fabian} A.~C., {Cackett} E.~M., {Uttley} P., {Wilkins} D.~R.,
  {Zoghbi} A., 2013{\natexlab{b}}, \mnras

\bibitem[{{Middleton}, {Done} \& {Gierli{\'n}ski}(2007){Middleton}, {Done}, \&
  {Gierli{\'n}ski}}]{Middleton07}
{Middleton} M., {Done} C., {Gierli{\'n}ski} M., 2007, \mnras, 381, 1426

\bibitem[{{Miller}, {Turner} \& {Reeves}(2008){Miller}, {Turner}, \&
  {Reeves}}]{Miller08}
{Miller} L., {Turner} T.~J., {Reeves} J.~N., 2008, \aap, 483, 437

\bibitem[{{Nardini} {et~al}\mbox{.}(2015){Nardini}, {Reeves}, {Gofford},
  {Harrison}, {Risaliti}, {Braito}, {Costa}, {Matzeu}, {Walton}, {Behar},
  {Boggs}, {Christensen}, {Craig}, {Hailey}, {Matt}, {Miller}, {O'Brien},
  {Stern}, {Turner}, \& {Ward}}]{Nardini15}
{Nardini} E. {et~al.}, 2015, Science, 347, 860

\bibitem[{{Parker} {et~al}\mbox{.}(2015){Parker}, {Fabian}, {Matt}, {Koljonen},
  {Kara}, {Alston}, {Walton}, {Marinucci}, {Brenneman}, \&
  {Risaliti}}]{Parker15_pcasample}
{Parker} M.~L. {et~al.}, 2015, \mnras, 447, 72

\bibitem[{{Parker} {et~al}\mbox{.}(2014{\natexlab{a}}){Parker}, {Marinucci},
  {Brenneman}, {Fabian}, {Kara}, {Matt}, \& {Walton}}]{Parker14_mcg6}
{Parker} M.~L., {Marinucci} A., {Brenneman} L., {Fabian} A.~C., {Kara} E.,
  {Matt} G., {Walton} D.~J., 2014{\natexlab{a}}, \mnras, 437, 721

\bibitem[{{Parker} {et~al}\mbox{.}(2017){Parker}, {Pinto}, {Fabian}, {Lohfink},
  {Buisson}, {Alston}, {Kara}, {Cackett}, {Chiang}, {Dauser}, {De Marco},
  {Gallo}, {Garcia}, {Harrison}, {King}, {Middleton}, {Miller}, {Miniutti},
  {Reynolds}, {Uttley}, {Vasudevan}, {Walton}, {Wilkins}, \&
  {Zoghbi}}]{Parker17_iras}
{Parker} M.~L. {et~al.}, 2017, \nat, 543, 83

\bibitem[{{Parker} {et~al}\mbox{.}(2014{\natexlab{b}}){Parker}, {Schartel},
  {Komossa}, {Grupe}, {Santos-Lle{\'o}}, {Fabian}, \&
  {Mathur}}]{Parker14_mrk1048}
{Parker} M.~L., {Schartel} N., {Komossa} S., {Grupe} D., {Santos-Lle{\'o}} M.,
  {Fabian} A.~C., {Mathur} S., 2014{\natexlab{b}}, \mnras, 445, 1039

\bibitem[{{Parker} {et~al}\mbox{.}(2014{\natexlab{c}}){Parker}, {Walton},
  {Fabian}, \& {Risaliti}}]{Parker14_ngc1365}
{Parker} M.~L., {Walton} D.~J., {Fabian} A.~C., {Risaliti} G.,
  2014{\natexlab{c}}, \mnras, 441, 1817

\bibitem[{{Tombesi} {et~al}\mbox{.}(2010){Tombesi}, {Cappi}, {Reeves},
  {Palumbo}, {Yaqoob}, {Braito}, \& {Dadina}}]{Tombesi10}
{Tombesi} F., {Cappi} M., {Reeves} J.~N., {Palumbo} G.~G.~C., {Yaqoob} T.,
  {Braito} V., {Dadina} M., 2010, \aap, 521, A57

\bibitem[{{Tombesi} {et~al}\mbox{.}(2015){Tombesi}, {Mel{\'e}ndez}, {Veilleux},
  {Reeves}, {Gonz{\'a}lez-Alfonso}, \& {Reynolds}}]{Tombesi15}
{Tombesi} F., {Mel{\'e}ndez} M., {Veilleux} S., {Reeves} J.~N.,
  {Gonz{\'a}lez-Alfonso} E., {Reynolds} C.~S., 2015, \nat, 519, 436

\bibitem[{{Uttley} {et~al}\mbox{.}(2014){Uttley}, {Cackett}, {Fabian}, {Kara},
  \& {Wilkins}}]{Uttley14}
{Uttley} P., {Cackett} E.~M., {Fabian} A.~C., {Kara} E., {Wilkins} D.~R., 2014,
  \aapr, 22, 72

\bibitem[{{Vaughan} {et~al}\mbox{.}(2003){Vaughan}, {Edelson}, {Warwick}, \&
  {Uttley}}]{Vaughan03_variability}
{Vaughan} S., {Edelson} R., {Warwick} R.~S., {Uttley} P., 2003, \mnras, 345,
  1271

\end{thebibliography}

\end{document}